\documentclass[journal= jcisd8,manuscript=article]{achemso}

\usepackage{chemformula} 
\usepackage[T1]{fontenc} 

\author{E.G.C. Matias}
\affiliation[UFRN]{Departamento de Biof\'isica e Farmacologia, Universidade Federal do Rio Grande do Norte, 59072-970, Natal-RN, Brazil.}

\author{K. S. Bezerra}
\affiliation[UFRN]{Departamento de Biof\'isica e Farmacologia, Universidade Federal do Rio Grande do Norte, 59072-970, Natal-RN, Brazil.}

\author{A.H. Lima Costa}
\affiliation[UFRN]{Departamento de Biof\'isica e Farmacologia, Universidade Federal do Rio Grande do Norte, 59072-970, Natal-RN, Brazil.}

\author{W. S. Clemente}
\affiliation[UFRN]{Departamento de F\'isica Te\'orica e Experimental, Universidade Federal do Rio Grande do Norte, 59072-970, Natal-RN, Brazil.}

\author{J. I. N. Oliveira}
\affiliation[UFRN]{Departamento de Biof\'isica e Farmacologia, Universidade Federal do Rio Grande do Norte, 59072-970, Natal-RN, Brazil.}

\author{L. A. Ribeiro Junior}
\affiliation[UNB]{Institute of Physics, University of Brasília, 70919-970, Bras\'lia-DF, Brazil.}

\author{D. S. Galvão}
\affiliation[UNICAMP]{Applied Physics Department, University of Campinas, Campinas, São Paulo, Brazil.}

\author{U. L. Fulco}
\affiliation[UFRN]{Departamento de Biof\'isica e Farmacologia, Universidade Federal do Rio Grande do Norte, 59072-970, Natal-RN, Brazil.}
\email{umbertofulco@gmail.com}
\phone{+-55 (84) 3215-3419}
\fax{+-55 (84) 3215-3791}

\title{Quantum Biochemical Analysis of the TtgR Regulator and Effectors}


\begin{document}

\begin{tocentry}

\includegraphics[width=5.0cm]{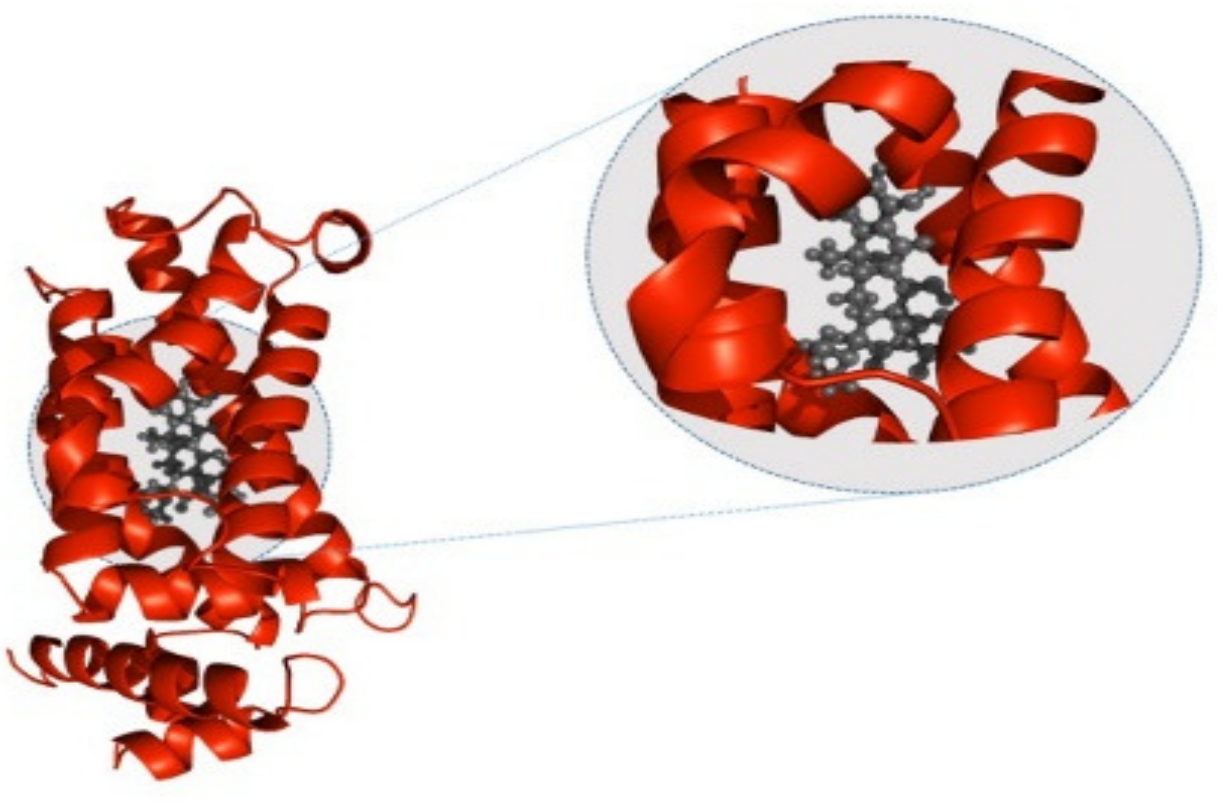}

\end{tocentry}

\begin{abstract}
The recent expansion of multidrug-resistant (MDR) pathogens poses significant challenges in treating healthcare-associated infections. Although antibacterial resistance occurs by numerous mechanisms, active efflux of the drugs is a critical concern. A single species of efflux pump can produce a simultaneous resistance to several drugs. One of the best-studied efflux pumps is the TtgABC: a tripartite resistance-nodulation-division (RND) efflux pump implicated in the intrinsic antibiotic resistance in \textit{Pseudomonas putida} DOT-T1E. The expression of the TtgABC gene is down-regulated by the HTH-type transcriptional repressor TtgR. In this context, by employing quantum chemistry methods based on the Density Functional Theory (DFT) within the Molecular Fragmentation with Conjugate Caps (MFCC) approach, we investigate the coupling profiles of the transcriptional regulator TtgR in complex with quercetin (QUE), a natural polyphenolic flavonoid, tetracycline (TAC), and chloramphenicol (CLM), two broad-spectrum antimicrobial agents. Our quantum biochemical computational results show the: [i] convergence radius, [ii] total binding energy, [iii] relevance (energetically) of the ligands regions, and [iv] most relevant amino acids residues of the TtgR-QUE/TAC/CLM complexes, pointing out distinctions and similarities among them. These findings improve the understanding of the binding mechanism of effectors and facilitate the development of new chemicals targeting TtgR, helping in the battle against the rise of resistance to antimicrobial drugs.
\end{abstract}

\section{Introduction}

Bacterial resistance is a critical public health problem worldwide, partly due to antibiotics overuse. The World Health Organization (WHO) listed antibiotic resistance as one of the top three threats to public health in the 21st century. Antibiotic resistance to common bacteria has already reached alarming levels in many parts of the world, compromising the medical treatments available for even common infections \cite{Munita2016, Kumar2012, WHO2014}.

Antimicrobial resistance is a consequence of the interaction of many organisms with their environment as a process of natural selection. Most antimicrobial compounds are naturally produced molecules, and, as such, co-resident bacteria have developed mechanisms to overcome their actions to survive \cite{Munita2016}. This phenomenon is associated with several factors, which involve mutations in common resistance genes, exchange of genetic information between microorganisms, development of environmental conditions in hospitals and communities, and the proliferation and dissemination of resistant bacterial clones \cite{Tenover2001}.

Bacteria have several ways to evade antimicrobial agents, including the production of inactivating enzymes, changes in the active sites, membrane impermeability, efflux, or the presence of enzyme resistance genes. One of the main mechanisms of bacterial resistance to antimicrobial agents includes the active efflux of drugs  \cite{Livermore2003, Kumar2012, Munita2016}. Efflux pumps are the most ubiquitous resistance element in all organisms, from bacteria to mammals. This mechanism is usually associated with resistance to multiple drugs (MDR). It provides active transport of harmful compounds from the membrane or cytoplasm to the external environment, reducing the concentration of antimicrobial in the intracellular space to sub-toxic levels \cite{Alguel2007, Blanco2016}.

Developing MDR resistance is a consequence of the presence of the same genetic mobile element of several genes. Each gene encodes a different resistance determinant. However, in some circumstances, the same determinant can confer resistance to different antimicrobials characterizing the MDR efflux pumps \cite{Blanco2016}. MDR pumps are associated with specific protein families --- the members of that family are the most relevant concerning resistance to agents of clinical importance --- being present in significant Gram-negative bacteria, such as \textit{P. aeruginosa}, \textit{E. coli}, \textit{Salmonella enterica}, \textit{Serovar Typhimurium}, \textit{Campylobacter jejuni}, and \textit{Neisseria gonorrhoeae} \cite{Teran2006, Piddock2006, Nikaido2009}.

Resistance-nodulation-division (RND) superfamily plays a crucial role in the production of MDR resistance \cite{Nikaido2009}. This feature is due to the transporters of this family that form a multi-component system associated with two other classes of proteins: the membrane fusion protein (MFP) and the outer membrane factor (OMF) \cite{Alguel2007, Diaz2003, Piddock2006, Sun2014}.

Patterns of this family are organized in a tripartite efflux mechanism, constituted, in most cases, by three components: a carrier protein present in the inner membrane (for example, AcrB), a periplasmic accessory protein (AcrA), which is the fusion protein membrane (MFP), and an outer membrane protein channel (TolC), which is the outer membrane factor (OMF). AcrB captures its substrates within the phospholipid bilayer of the inner membrane or cytoplasm and transports them to the external medium through TolC, external membrane factor (OMF) \cite{Piddock2006}.

The set formed by the three RND-MFP-OMF proteins composes a continuous channel through the envelope of Gram-negative bacteria. This channel ensures that the substrate molecule, captured from the inner membrane of the phospholipid bilayer, is effluxed directly through the periplasm and the outer membrane. Therefore, they work as anti-gates, using the proton gradient across the membrane for inducing the efflux, \cite{Piddock2006, Fernando2013}. The tripartite RND efflux systems are regulated and encoded by genes, which are generally organized as an operon. They correspond to the set formed by the promoter, the operator, and the regulatory genes, which are arranged side by side \cite{Diaz2003, Piddock2006}.

TtgR is a multi-drug binding repressor that negatively controls the transcription of the operon ttgABC-ttgR, an efflux pump present in the species \textit{P. putida} DOT-T1E \cite{Teran2006}. This bacterium is resistant to multiple antibiotics and capable of surviving in the presence of secondary plant metabolites. The responsible for this phenotype is the ttgABC- ttgR efflux pump \cite{Daniels2010}. TtgR repressor belongs to the TetR family of proteins have two DNA binding domains, an N-terminal domain and a C-terminal domain. Generally, these proteins have a helical shape and function as dimers \cite{Cuthbertson2013}. They are similar to QacR, a multi-drug binding protein of \textit{S. aureus} \cite{Alguel2007}.

The TtgR structure is based on monomers consisting of nine alpha-helices that form two distinct domains. The first one is the DNA binding (C-terminal) domain, which consists of {$\alpha$}1-{$\alpha$}3, with helix 3 forming most of the contacts with DNA. The second corresponds to the ligand-binding domain that encompasses {$\alpha$}4-{$\alpha$}9, helix 4. It serves as a junction for the rest of the protein, which folds independently of the N-terminal domain, and constitutes the effector binding pocket \cite{Daniels2010, Alguel2007}.

TtgR performs its function through an effector-mediated deregulation mechanism. In the absence of effectors (antibiotics, flavonoids, and others), TtgR is naturally binding to the region of the gene operator, repressing the transcription of the TtgABC gene. The association of effectors to the TtgR-DNA complex leads to the dissociation of TtgR from the operator region, allowing transcription. TtgR can bind to multiple substances, one of the characteristics of the proteins involved in the regulation of MDR pumps \cite{Krell2007, Teran2006}. The effectors (or ligands) can be vertically coupled to TtgR through a general/hydrophobic binding pocket with few specific interactions. They can also bind to a second binding pocket, known as the high-affinity pocket or specific binding \cite{Alguel2007}.

Based on the clinical importance of the multicomponent resistance mechanism RND-MFP-OMF that constitutes the TtgABC efflux system, it is necessary to develop studies on the effector binding properties to the inhibitor of TtgR. From its structural analysis, it is possible to delineate the most relevant interactions for the coupling mechanism of its effectors. This analysis can pave the way for developing new efficient compounds to overcome this resistance mechanism. 

In the present work, we propose a \textit{in silico} structural, biochemical analysis of the protein and transcriptional regulator TtgR of the TtgABC efflux pump in a complex with three effectors: quercetin (a polyphenolic bioflavonoid found in fruits and vegetables with antibacterial properties \cite{Jaisinghani2017}), and two broad-spectrum antimicrobial agents, the tetracycline and chloramphenicol \cite{Chopra2001}. The main interactions between the TtgR protein and quercetin (QUE), tetracycline (TAC), and chloramphenicol (CLM) were described from their crystallographic data. Through computational biochemistry methods, interaction energy calculations were carried out within the Functional Density Theory (DFT) formalism by employing a Molecular Fragmentation with Conjugate Caps (MFCC) approach. This approach has been successfully used for proposing new pharmacological perspectives \cite{Cui2016, Santos2022, Barbosa2021, Katy2019, Katy2020, LimaCosta2018, Tavares2019}. 

\section{Materials and methods}
To carry out the computational study for the intermolecular interactions of the TtgR protein with the effectors, energetic calculations considered the cryptographic structures available in the PDB database (http://www.rcsb.org), namely: 2UXH - 2.4 {\AA}, 2UXO - 2.7 {\AA} and 2UXP (2.7 {\AA}) for quercetin, tetracycline, and chloramphenicol in complex with TtgR, respectively \cite{Alguel2007}. Initially, atoms not resolved by X-ray diffraction (as hydrogens and some amino acid side chains) were added to the structures. The protonation state of the receptor TtgR and its effectors at physiological pH were adjusted according to the results obtained from the MarvinSketch code version 17.24 (Marvin Beans Suite - ChemAxon) and the PROPKA 3.1  \cite{PROPKA2011} package, respectively. Aminoacid main-chain atoms were maintained fixed while side-chains were submitted to a classical geometry optimization using the classic CHARMM (Chemistry at Harvard Molecular Mechanics) force field \cite{Momany1992}. The calculations were carried out with convergence tolerances set to 10$^{-5}$ kcal/mol (total energy change), 10$^{-3}$ kcal/mol (mean square root of the RMS gradient) and 10$^{-5}$ {\AA} (maximum atomic displacement).

The binding energy between the effectors and the TtgR protein was calculated using the MFCC scheme \cite{albuquerque2021quantum, vianna2021new, Albuquerque2014}. This approach has been widely used for the energetic analysis of protein-ligand and protein-protein complexes \cite{Katy2018, Xavier2017, Katy2019, Zhang2017, Brinkmann2014, Wang2013, Dantas2015, Tavares2018, Katy2017}. It allows evaluating a considerable number of amino acids in a given macromolecule with a relatively small computational expenditure, which is crucial in studying complex biological systems.

In the framework of this approach, a protein is decomposed into individual amino acid fragments by cutting through the peptide bond. To preserve the local chemical environment and comply with the valence requirements, a pair of conjugate caps is designed to saturate each fragment, and hydrogen atoms are added to the molecular caps to avoid dangling bonds \cite{Gordon2012, Zhang2003}. Therefore, the energy of interaction between a protein and an effector can be achieved through arrangements of interaction energies between individual fragments of the analyzed system, also providing an interpretation of the local environment during individual fragmentation.

Here, we label the ligands/effectors as $ L $ and the residue that interacts with $ L $ as $ R_ {i} $, characterizing the \textit{i}-th amino acid residue. The cap $ C_ {i-1} (C_ {i + 1}) $ is formed by the neighbor residue covalently bound to the amine (carboxyl) group residue $ R_ {i} $ along the protein chain, providing a better characterization of the electronic component environment. For these fragmented structures, the energy of the interaction between the effector and the individual fragments, $ E (L-R_ {i}) $, is calculated according to:

\begin{eqnarray} \label{eq1}
E(L-R_{i})&=& E(L- C_{i-1}R_{i}C_{i+1}) - E(L- C_{i-1}C_{i+1}) \nonumber \\
&-& E(C_{i-1}R_{i}C_{i+1}) + E(C_{i-1}C_{i+1}),
\end{eqnarray}

\noindent where the first term, $E(L- C_{i-1}R_{i}C_{i+1})$, is the total energy of the system formed by the ligand and the capped residue; the second term, $E(L- C_{i-1}C_{i+1})$, is the total energy of the system formed by the caps and the ligand. Also, the third term, $E(C_{i-1}R_{i}C_{i+1})$, is the total energy of the residue with the caps, while the fourth and final one, $E(C_{i-1}C_{i+1})$, is the energy of the caps with dangling bonds hydrogenated.

After fragmentation, energetic calculations were performed for each ligand–residue intermolecular interaction at the binding site using Gaussian 09W software \cite{Gaussian09}, within the DFT formalism. We considered the generalized gradient approximation (GGA) and the B97D functional  \cite{Grimme2006}. This functional includes a dispersion term (D2) with a damping function configured with a fixed value of 6.0. It shows good performance for non-covalently bound systems, being a quantum efficient and accurate chemical method for large systems, where the magnitude of the dispersion forces are relevant \cite{Li2014}. To represent the electronic wave function and to expand the Kohn-Sham orbitals for all electrons, we selected the 6-311 + G (d, p), a triple split valence (valence triple-zeta) small basis set along with an additional diffuse function (+) and polarization functions (\textit{d, p}).

To represent the molecular environment, we used the Continuous Conductive Polarizable Model (CPCM) \cite{Cossi2003} with dielectric constants ($\varepsilon$) equal to 10 and 40. This approach was considered to increase the similarity with the protein environment and estimate energetic effects, such as the electrostatic polarization promoted by the solvent.

We carried out a convergence study for the total binding energy as a function of the radius of the binding pocket of the ligand $r$ from 2.0 to 13.0 {\AA}. This radius limits the number of residues of amino acids to be analyzed. In this sense, one can mitigate the loss of relevant interactions. Fictitious spheres were employed (considering $r = n/2; n = 1, 2, 3, 4, ...$) for demarcating increasing distances from the effectors to determine the sum of the individual energies of the residues present in each one of them. Thus, convergence is achieved when the energy change in the consecutive radius is less than 10\% \cite{Vianna2019, Xavier2020}.

\section{Results and Discussions}

\begin{figure}[!htb]
\vspace{0.0cm}
\centering
\includegraphics[width=0.5\textwidth,keepaspectratio]{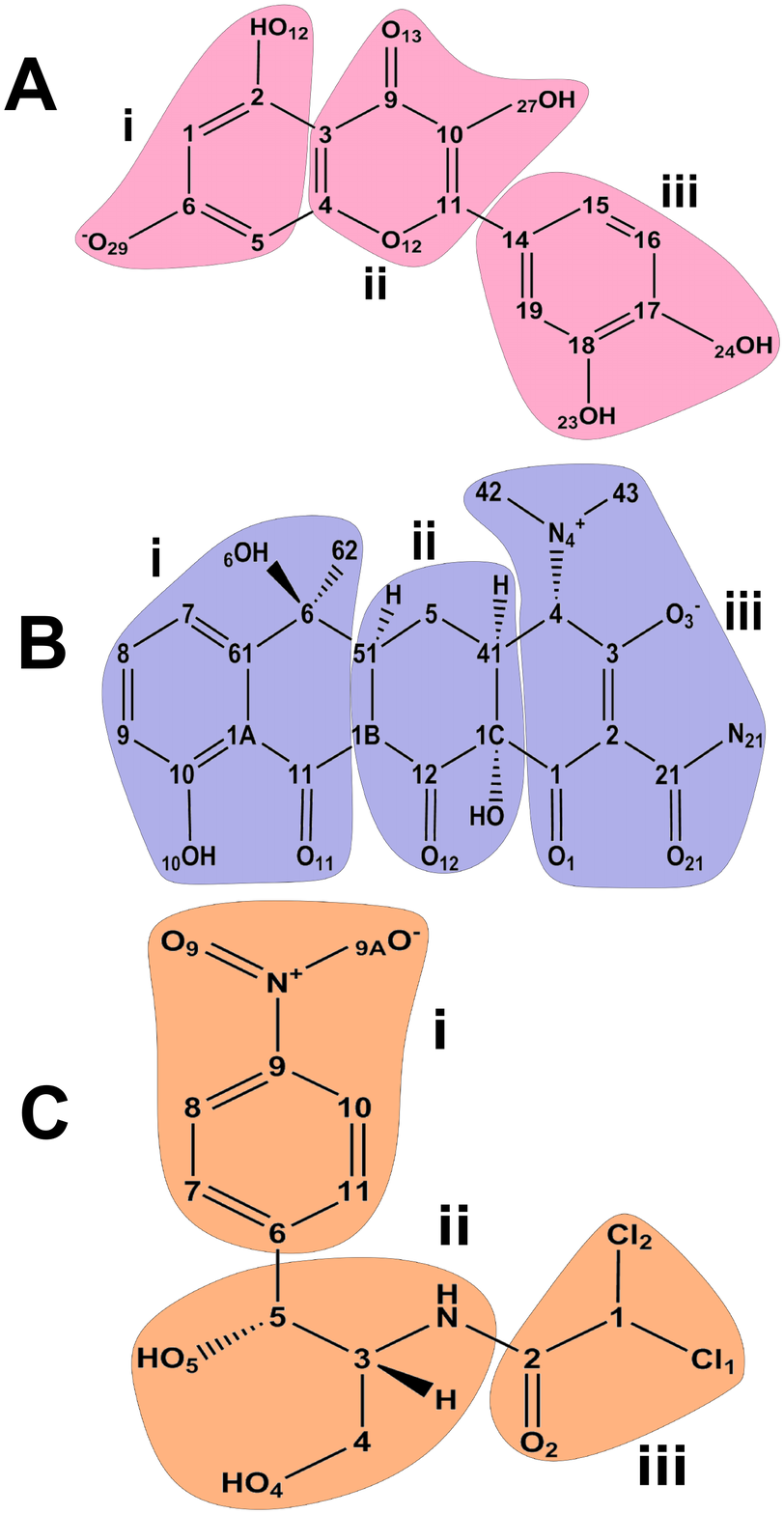}
\caption{Schematic representation of the chemical structure of the effectors (a)  QUE, (b) TAC, and (c) CLM, subdivided into three regions i, ii, iii.}
\vspace{0.0cm}
\label{Fig1}
\end{figure}

To better expose the interactions that occur in each effector with the TtgR protein, we schematically divide the three ligands into regions. The three effectors QUE, TAC, and CLM were subdivided into three regions, i, ii, and iii, as shown in Fig.\ref{Fig1}.

Interaction energies and convergence criteria were evaluated for each of the three ligands. They were duly defined by the radius ($ r $) of the  bond pocket (in {\AA}) and its related interaction energy (in kcal/mol). For the analysis of the binding interactions between the complexes, it was necessary to observe all the relevant attractive and repulsive amino acid residues. They may interfere in the systems for the development of new therapeutic agents

The total binding energy for each radius in the binding pocket (local) was obtained from the sum of the individual energies of each amino acid residue, evaluating the convergence. Concerning the attractive interaction energy (negative energy) or repulsion (positive energy) for each relevant amino acid residue, an analysis was performed considering an ideal radius in the binding pockte, for which there would be no significant variation (less than 10 \%) of the total interaction energy.

\begin{figure}[htb]
\centering
\vspace{0.0cm}
\includegraphics[width=0.45\textwidth,keepaspectratio]{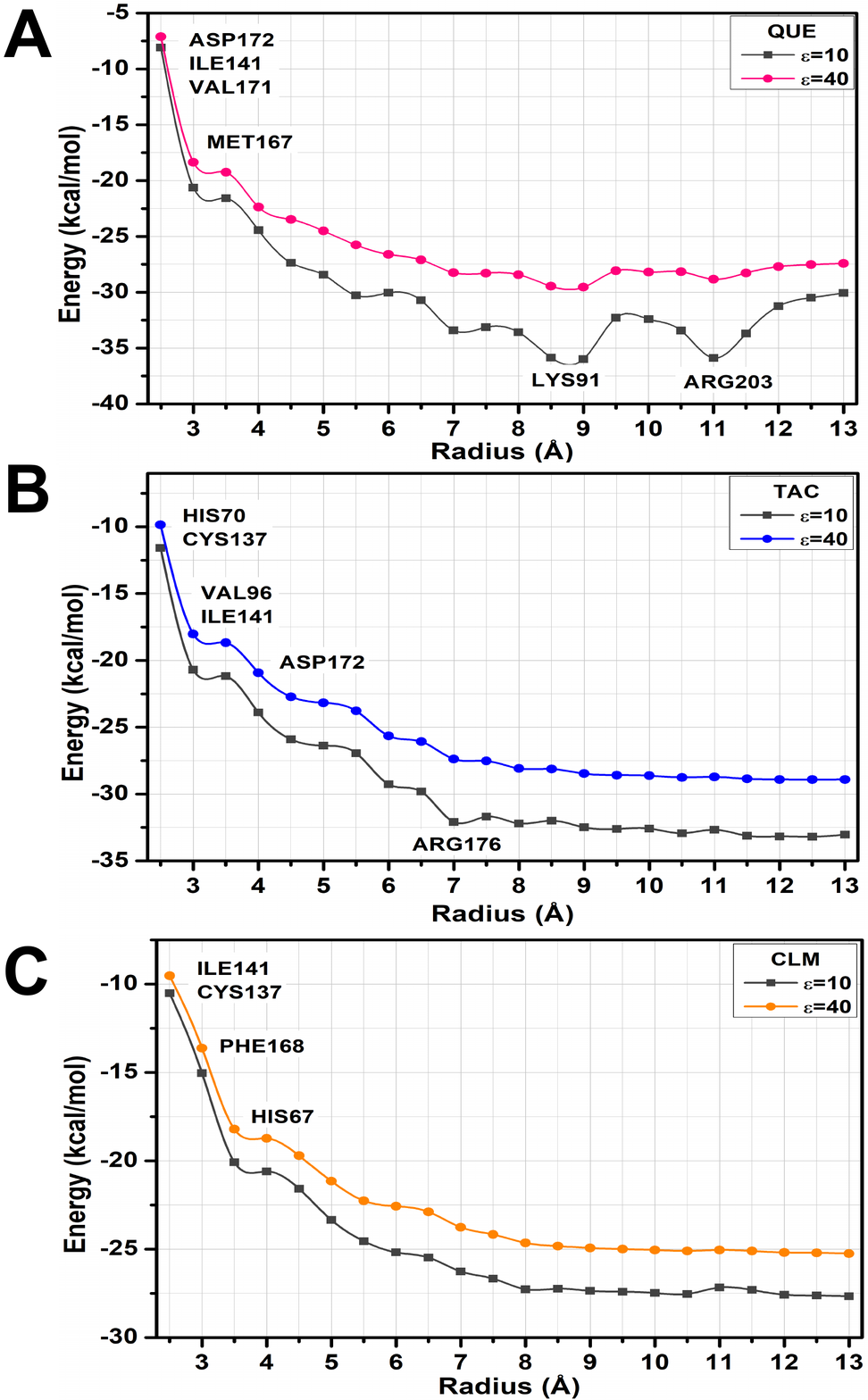}
\caption{Total interaction energy (in kcal/mol) of the TtgR complex ligands as a function of the binding pocket radius  $r$ (in {\AA}) was calculated using the functional GGA B97D within the MFCC scheme. Two dielectric constants are: $\varepsilon $= 10 and 40. The highlighted amino acid residues are those responsible for the largest binding energies.}
\centering
\vspace{0.0cm}
\label{Fig2}
\end{figure}

Figure \ref{Fig2} represents the calculated binding energy of the complex TtgR interacting with the ligands QUE, TAC, and CLM as a function of the radius $ r $, considering the constant dielectrics $\varepsilon $ = 10 and 40. From this figure, it can be seen that the convergence occurs from the radius $ r = $ 9.0, 7.5, and 6.5 {\AA} (considering $\varepsilon $ = 40) for QUE, TAC, and CLM, respectively. To estimate the contribution of a considerable number of residues and to guarantee the evaluation of all important residues for the interactions between the systems, we performed all calculations for radius values up to 13 {\AA}, comprising a total of 122 interactions (QUE), 133 (TAC), and 111 (CLM) for the complexes. The total energy for each system reached the values of -30.06 kcal/mol ($\varepsilon$=10) and -27.42 kcal/mol ($\varepsilon$=40) for QUE, -33.16 kcal/ mol ($\varepsilon$=10) and -28.79 kcal/mol ($\varepsilon$=40) for TAC, and -27.67 kcal/mol ($\varepsilon$=10) and -25.25 kcal/ mol ($\varepsilon$=40) for CLM.

Considering the residues evaluated for the TtgR-QUE complex, eleven of them were listed as main. Only two amino acids have repulsive interactions (positive energies). Therefore, we have (for $\varepsilon$=10; $\varepsilon$=40): ASP172 (6.0; 4.23) and GLU78 (3.42; 0.79). The rest (nine residues) presented attractive interactions (negative energies). In this sense, we obtained (for $\varepsilon$=10; $\varepsilon$=40): PHE168 (-4.72; -4.65), VAL171 (-4.53; -3.92), ILE141 (-4.35; -2.83), MET167 (-3.33; -2.51), LEU93 (-3.16; -2.82), ARG176 (-2.54; -0.85), LYS91 (-2.45; -0.66), VAL96 (-2.38; -2.09), and LEU92 (-2.09; -1.78), in kcal/mol.

\begin{figure}[htb!]
\centering
\vspace{0.0cm}
\includegraphics[width=0.5\textwidth,keepaspectratio]{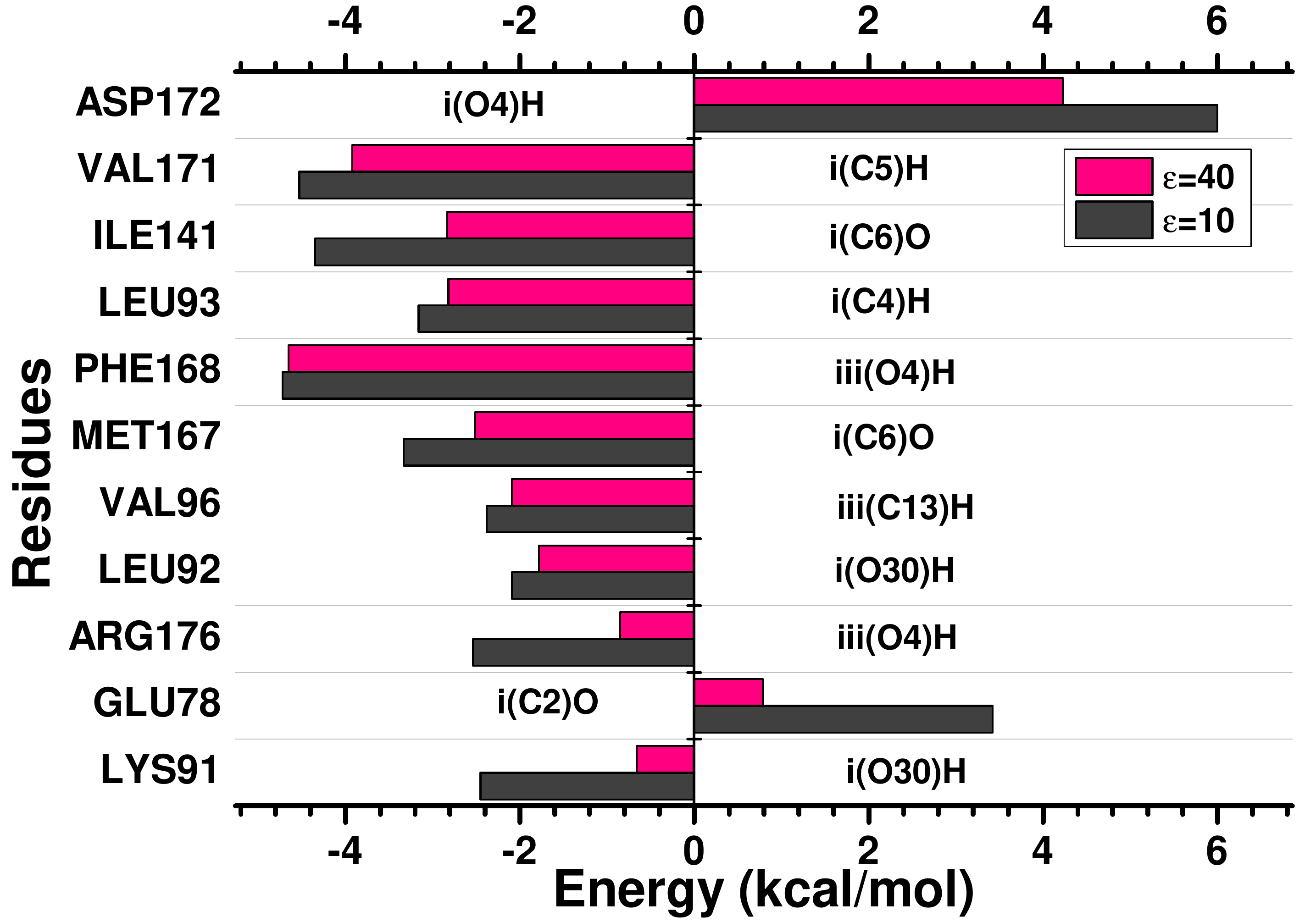}
\caption{Graphic panels show the most relevant residues regarding the contribution of the interaction energies of the TtgR-QUE biological complex system. The regions and atoms of ligands that interact with each residue at the binding site are also shown.}
\centering
\vspace{0.0cm}
\label{Fig3}
\end{figure}

Figure \ref{Fig3} shows the energy values obtained for the residues listed as the most important for the TtgR-QUE binding site and the distances (in {\AA}) between residue-ligand. The interaction energies by region are -16.18 kcal/mol ($\varepsilon$=10) and -14.04 kcal/mol ($\varepsilon$=40) for region i, and -3.64 kcal/mol ($\varepsilon$=10) and -3.36 kcal/mol ($\varepsilon$=40) for region iii. Region i is characterized as the most energetic. This trend occurs once it presents a higher amount of residues with attractive interactions due to the presence of an oxygen atom (O29), which contains a negative charge making the region apt to attract neutral or positively charged residues. On the other hand, the reduction in energy in region iii results from a lower number of interactions, in addition to the high repulsive value achieved by ASP172.

The residue that presented the highest attractive interaction energy was PHE168 in the radius of 3.0 {\AA}, reaching the values of -4.72 kcal/mol ($\varepsilon$=10) and -4.65 kcal/mol ($\varepsilon$=40). The inretaction occurs in region iii (O4) H from dipole-dipole forces and has non-conventional hydrogen bonds. Dipole-dipole interactions result from the communication between two groups with a polarization of opposite charges due to the difference in electronegativity of the atoms involved \cite{Sippel2015}. This residue is part of the hydrophobic environment of the sidewalls of the {$\alpha$}8 helix, interacting with the quercetin chromenone rings, which are close to the top of the binding site.

PHE168 is one of the amino acids in the general bind pocket shared by the high-affinity site \cite{Alguel2007}. The second-highest value of attractive energy was reached by the amino acid VAL171 in the radius of 2.5 {\AA}, reaching the values of -4.36 kcal/mol ($\varepsilon$=10) and -3.92 kcal/mol ($\varepsilon$=40). The interactions occurred in regions i(C5)H and ii(C11)H using London forces and alkyl-$pi$ interactions. These types of interactions are considered weak, but they are relevant in the molecular recognition process of the drug at its binding site. They generally form multiple interactions, which make them significant energy contributions \cite{Sippel2015}. Like PHE168, this residue provides a hydrophobic environment on the sidewalls of the {$\alpha$}8 helix communicating with the QUE hydroxyphenyl ring \cite{Alguel2007}.

The amino acid ILE141 (Fig.\ 4) interacts with QUE through region i(C6) and i(C1)H in the 2.5 {\AA} radius, reaching the values of -4.35 kcal/mol ($\varepsilon$=10) and -2.83 kcal/mol ($\varepsilon$=40). The binding occurred through non-conventional H bond interactions and London forces. This residue is located in the {$\alpha$}7 helix, and is part of the hydrophobic environment, being located near the top of the connection pocket, if communicated with QUE chromenone rings \cite{Alguel2007}. 

\begin{figure}[htb!]
\centering
\vspace{0.0cm}
\includegraphics[width=0.5\textwidth,keepaspectratio]{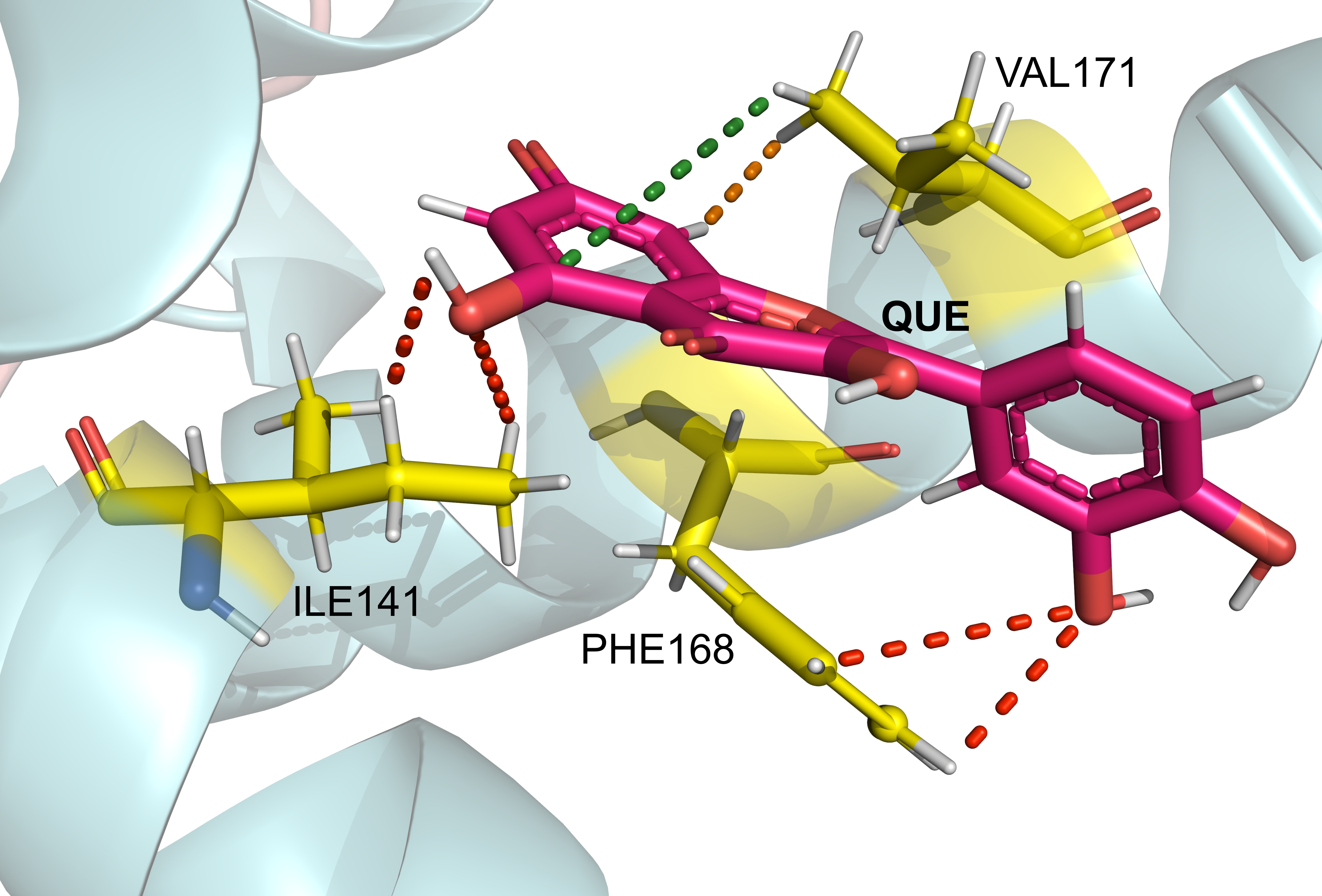}
\caption{Some of the main interactions for the TtgR-QUE complex. Dashed lines correspond to non-conventional H bonds (red), London forces (orange), and alkyl-$pi$ (green) interactions.}
\centering
\vspace{0.0cm}
\label{Fig4}
\end{figure}

MET167 presented the attractive energy values of -3.33 kcal/mol ($\varepsilon$=10) and -2.51 kcal/mol ($\varepsilon$=40) in the 3.0 {\AA} radius, interacting with quercetin through the region i(C6)O and i(C5)O through dipole-dipole-induced connections. It is located in the {$\alpha$}8 helix and contributes to the binding site's hydrophobic environment. LEU93 (Fig.\ 4) interacts with QUE in the 2.5 {\AA} radius through London forces occurring in the effector region i(C4)H and i(C5)H, reaching values of -3.16 kcal/mol ($\varepsilon$=10) and -2.82 kcal/mol ($\varepsilon$=40). It is positioned on the {$\alpha$}5 helix, favoring the hydrophobic environment of the sidewalls with the residues mentioned earlier. Figure \ref{Fig4} shows the interactions between the amino acids VAL171, ILE141, and LEU93 with the QUE ligand. 

ARG176 is present in the {$\alpha$}8 helix. It is one of the polar residues that constitute the connection pocket and presents attractive energies of -2.54 kcal/mol ($\varepsilon$=10) and -0.85 kcal/mol ($\varepsilon$=40) in the radius of 5.0 {\AA}, interacting with the QUE in region iii(O4) H through dipole-dipole interactions. It is a unique residue of the strain DOT-T1E. In other species of Pseudomonas, the amino acid of this position is glycine. A study has shown that ARG176 is a crucial residue in the interaction at the high-affinity site with the florentine flavonoid \cite{Daniels2010}. Here, it proved relevant interacting with QUE, connecting with its hydroxyl group.

The amino acid LYS91 is located in the {$\alpha$}5 helix and reached the attractive energy values of -2.45 kcal/mol ($\varepsilon$=10) and -0.66 kcal/mol ($\varepsilon$=40) in a large radius, of 8.5 {\AA}, communicating with QUE using dipole-dipole forces in region i (O30) H. This attraction is explained by the chemical characteristic of the residue, basic polar, so it is positively charged and binds in a region of the effector close to the negative charge. VAL96 exhibited energy values of -2.38 kcal/mol ($\varepsilon$=10) and -2.09 ($\varepsilon$=40) for a radius of 3.0 {\AA}, interacting with the effector through London forces in region iii(C19)H and dipole-dipole in region ii(O27)H. Like LYS91, it is also present in the {$\alpha$}5 helix, contributing to the hydrophobic environment of the site, interacting with QUE's hydroxyphenyl ring \cite{Alguel2007}.

LEU92, like most of the attractive energy residues mentioned here, participates in the hydrophobic space of the connection pocket. It presented the energy values of -2.09 ($\varepsilon$=10) and -1.78 kcal/mol ($\varepsilon$=40) in the radius of 3.0 {\AA}, communicating with QUE through region i (O30) H using dipole-dipole forces and ii (C9) O13 by dipole-dipole-induced bonds. It is located on the {$\alpha$}5 helix, around the chromenone rings, similar to PHE168, ILE141, and LEU93.

The amino acids GLU78 and ASP172 showed the most relevant values of repulsive energy. GLU78 exposed a repulsive energy of 3.42 kcal/mol ($\varepsilon$=10) and 0.79 kcal/mol ($\varepsilon$=40) in a 6.0 {\AA} radius, interacting with the effector by means of dipole-dipole forces in region i(O30)H and dipole-dipole-induced bonds with i(C2)O30. It is positioned on the {$\alpha$}4 propeller. The study by Fernandez-Escamilla and collaborators \cite{Fernandez2015} showed a mutation from GLU78 to an alanine, E78A, with the TtgR protein and attested that this modification affects the stability of the protein when effector binding occurs at the binding site. Furthermore, it certifies that its location in {$\alpha$}4, part of the effector portal, is important in the ligand binding process and protein stability.

ASP172 was the residue that exhibited the most relevant repulsive force, presenting the values of 6.0 kcal/mol ($\varepsilon$=10) and 4.23 kcal/mol ($\varepsilon$=40) in the interaction radius of 2.0 {\AA}. This amino acid interacts with the effector through hydrogen bonds through region iii(O4)H and by dipole-dipole-induced forces [iii(C16)O4].

\begin{figure}[htb!]
\centering
\vspace{0.0cm}
\includegraphics[width=0.5\textwidth,keepaspectratio]{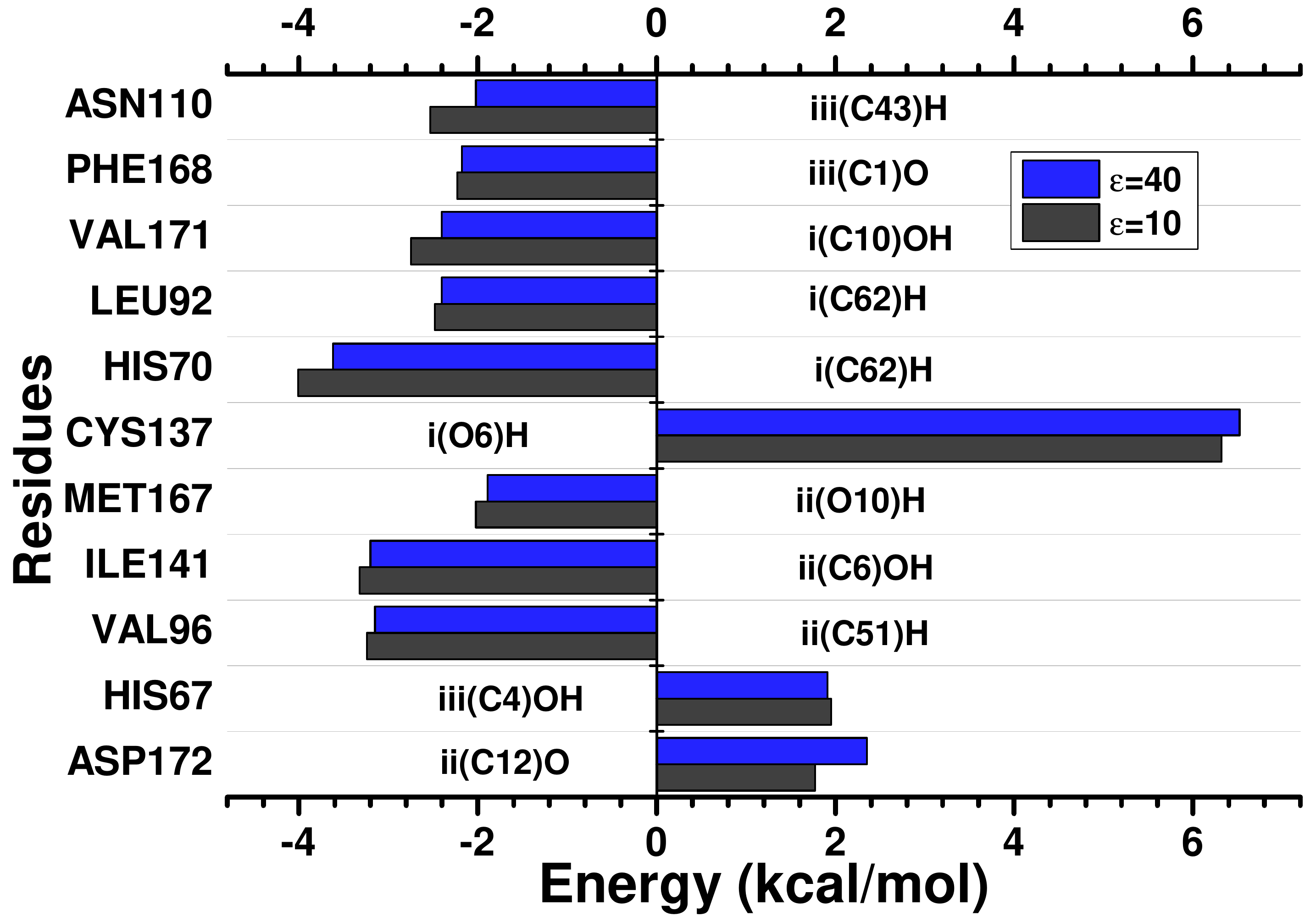}
\caption{The most relevant residues regarding the contribution of the interaction energies of the TtgR-TAC biological complex system. The regions and atoms of the ligands that interact with each residue at the binding site are also presented.}
\centering
\vspace{0.0cm}
\label{Fig5}
\end{figure}

The analysis of interactions between the TtgR-TAC complex, observed in eleven main amino acid residues, where three showed repulsive interactions, CYS137 ($\varepsilon$=10; $\varepsilon$=40)(6.52 kcal/mol; 6.32 kcal/mol), HIS67 (1.91 kcal/mol; 1.95 kcal/mol) and ASP172 (2.35 kcal/mol; 1.77 kcal/mol). The other eight residues presented attractive energies, where in decreasing order we have ($\varepsilon$=10; $\varepsilon$=40), in kcal/mol: HIS70 (-4.01; -3.62), ILE141 (-3.32; -3.20), VAL96 (-3.24; -3.15), VAL171 (-2.75; -2.40), LEU92 (-2.48; -2.40), ASN110 (-2.53; -2.02), ILE141 (-2.44; -2.31), PHE168 (-2.23; -2.18), MET167 (-2.02; -1.89), HIS67 (-1.95; -1.91). Figure \ref{Fig5} shows the energy values reached by the residue evaluated as the most important for the TtgR-TAC binding site, as well as the distance in angstroms ({\AA}) between the residue-effector.

Of the amino acids cited as relevant, three interact in the TAC i region (LEU92, LEU93, and ILE141), two binds in region ii (VAL96 and ARG176), and six (HIS70, VAL171, ARG130, LEU113, ASP64, and GLU111) communicate with the region iii of the TAC. The interaction energies by region for the TtgR-TAC complex were: region i was -10.40 kcal/mol ($\varepsilon$=10) and -9.92 kcal/mol ($\varepsilon$=40); for region ii, -7.02 kcal/mol ($\varepsilon$=10) and -7.07 kcal/mol ($\varepsilon$=40); and for region iii, -15.20 kcal/mol ($\varepsilon$=10) and -11.80 kcal/mol ($\varepsilon$=40). In region iii, the highest energy value was observed as it presents a higher amount of interactions regarding other places and shows those that reached the highest energy values.

HIS70 exhibited the highest attractive interaction energy for the TtgR-TAC complex, with interaction energies of -4.01 kcal/mol ($\varepsilon$=10) and -3.62 kcal/mol ($\varepsilon$=40) in the radius of 2.5 {\AA}. Alkyl-$pi$ i(C62)H and unconventional H bonds iii(C42)H interactions were observed. This residue, of basic polar characteristic, is located in the {$\alpha$}4 helix, where it forms a curve in the central region of this helix. This trend probably makes the entry or binding of effector molecules to TtgR \cite{Daniels2010} important. The study evaluated that mutation from HIS70 to an ALA, H70A, in the TtgR protein binding pocket affects the ability of antibiotics to access the overall binding pocket or prevents signal transduction of the binding response to the binding domain. DNA. It was also shown that this mutation does not affect the coupling of effectors that do not have \cite{Daniels2010} charges.

The hydrophobic residue VAL96 exhibited attractive interaction energy for this complex, -3.24 kcal/mol ($\varepsilon$=10) and -3.15 kcal/mol ($\varepsilon$=40), interacting within a radius of 2.5 {\AA} with the ligand regions iii(C42)H and ii(C51)H through London dispersion forces. Furthermore, VAL96 is relevant for interaction with TtgR since it is part of the {$\alpha$}4 helix that forms the sidewall of the connection pocket.  Moreover, this residue is one of those that interact with the cyclic rings of TAC \cite{Alguel2007}.

The amino acid ILE141 interacted with tetracycline within a radius of 3.0 {\AA}, with energy values of -3.32 kcal/mol ($\varepsilon$=10) and -3.20 kcal/mol ($\varepsilon$=40). Alkyl-$pi$ [i(C8)H] and unconventional H bonds [i(C6)OH] interactions are observed. This residue is positioned in the {$\alpha$}7 helix and interacts with the cyclic rings of the TAC along with other residues, participating in the hydrophobic environment of that \cite{Alguel2007} site. The LEU92 residue, on the other hand, presented energies of -2.48 kcal/mol ($\varepsilon$=10), -2.40 kcal/mol ($\varepsilon$=40), connecting to the effector in region i by forces of London [i(C7)H] and alkyl-alkyl [i(C62)H]. LEU92 is present in the {$\alpha$}5 helix near the top of the binding pocket. Like VAL96, VAL171, and ILE141, they also contribute to the hydrophobic environment of the sidewalls at the binding site, being around the cyclic rings of TAC \cite{Alguel2007, Daniels2010}.

Meanwhile, VAL171 interacted within a radius of 2.5 {\AA} with regions i and ii through unconventional H bonds [ii(C12)O, i(O11)O] with the TAC effector, reaching an energy of -2.75 kcal/mol ($\varepsilon$= 10) and -2.40 kcal/mol ($\varepsilon$=40). This amino acid is part of the {$\alpha$}8 helix and also makes up the hydrophobic environment of the side walls of this helix in TtgR \cite{Alguel2007}.

PHE168 showed interaction energy values of -2.23 kcal/mol ($\varepsilon$=10) and -2.18 kcal/mol ($\varepsilon$=40) at radius 2.5 {\AA}. The interaction occurs through unconventional H bonds in the [iii(C1)O, i(C11)O] region. This residue forms, together with LEU92, LEU93, VAL96, and VAL171, the sidewalls of the TtgR binding pocket, which, in turn, surround the cyclic tetracycline rings \cite{Alguel2007}. 

The amino acid ASN110 showed an H binding [iii(C3)O-] and an unconventional H binding [iii(C43)H]. This residue belongs to the bottom of the protein binding pocket, which is also composed of residues HIS114 and ASP172. Furthermore, ASN110 is positioned in a way that coordinates the interaction with the dimethylamine and amino groups of the effector \cite{Alguel2007}. The mutation resulting from the exchange of an ASN for an ALA residue at position 110 of TtgR produced a different dissociation scheme in the target operator. This fact denotes the importance of this residue in the mechanism of repression of the TtgR protein. \cite{Fernandez2015}. Likewise, MET167 interacts with TAC from radius 3.0 {\AA}, denoting London dispersion forces [i(C9)H] and unconventional H bonds [i(C10)H] bonds, presenting energies of -2.02 kcal/mol and -1.89 kcal/mol for $\varepsilon$=10 $\varepsilon$=40, respectively. 

\begin{figure}[htb!]
\centering
\vspace{0.0cm}
\includegraphics[width=0.5\textwidth,keepaspectratio]{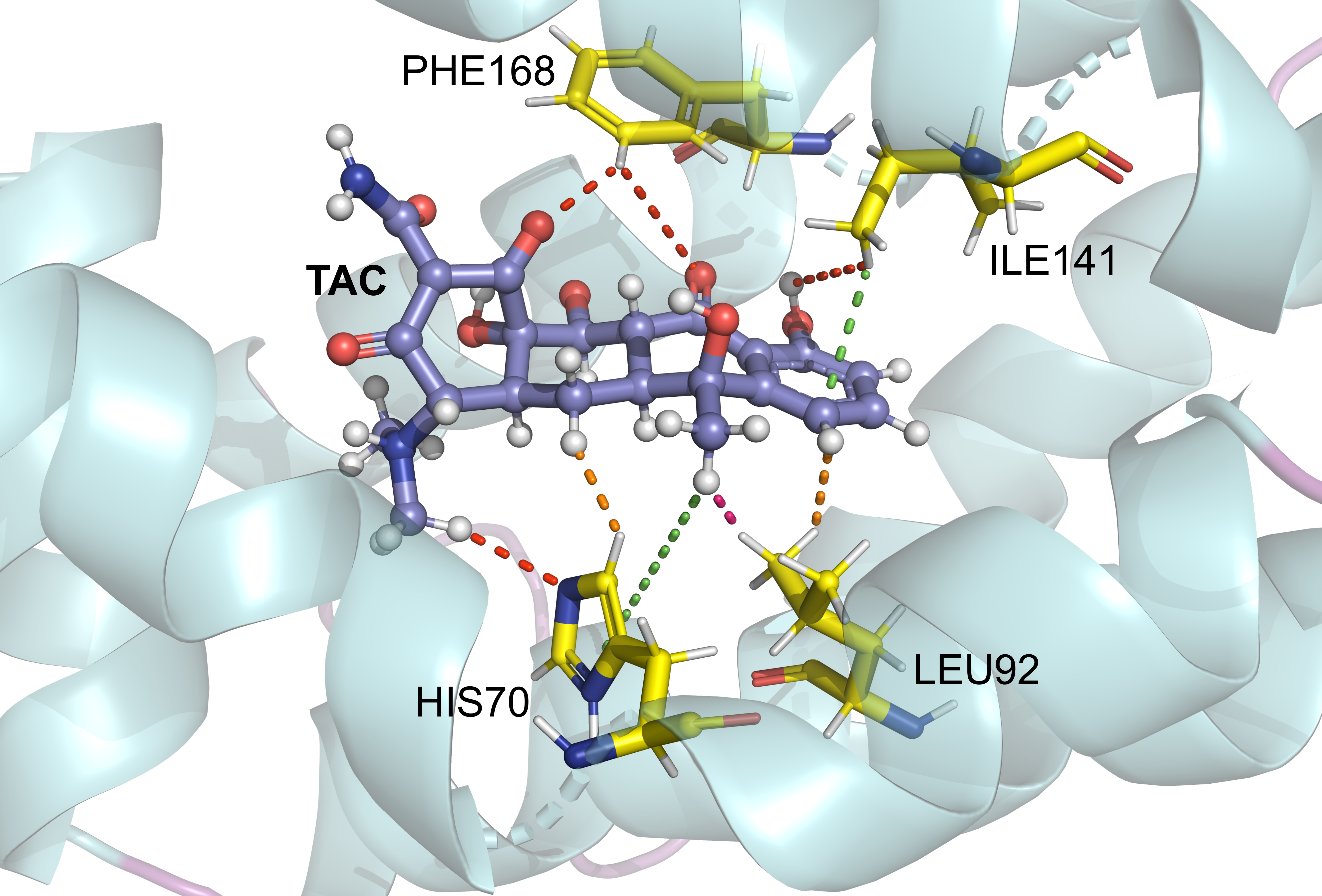}
\caption{Schematic representation of the main interactions for the TtgR-QUE complex. Dashed lines correspond to non-conventional H bonds (red), London forces(orange), alkil-$pi$ (green), and alkyl-alkyl (pink) interactions.}
\centering
\vspace{0.0cm}
\label{Fig6}
\end{figure}

\begin{figure}[htb!]
\centering
\vspace{0.0cm}
\includegraphics[width=0.5\textwidth,keepaspectratio]{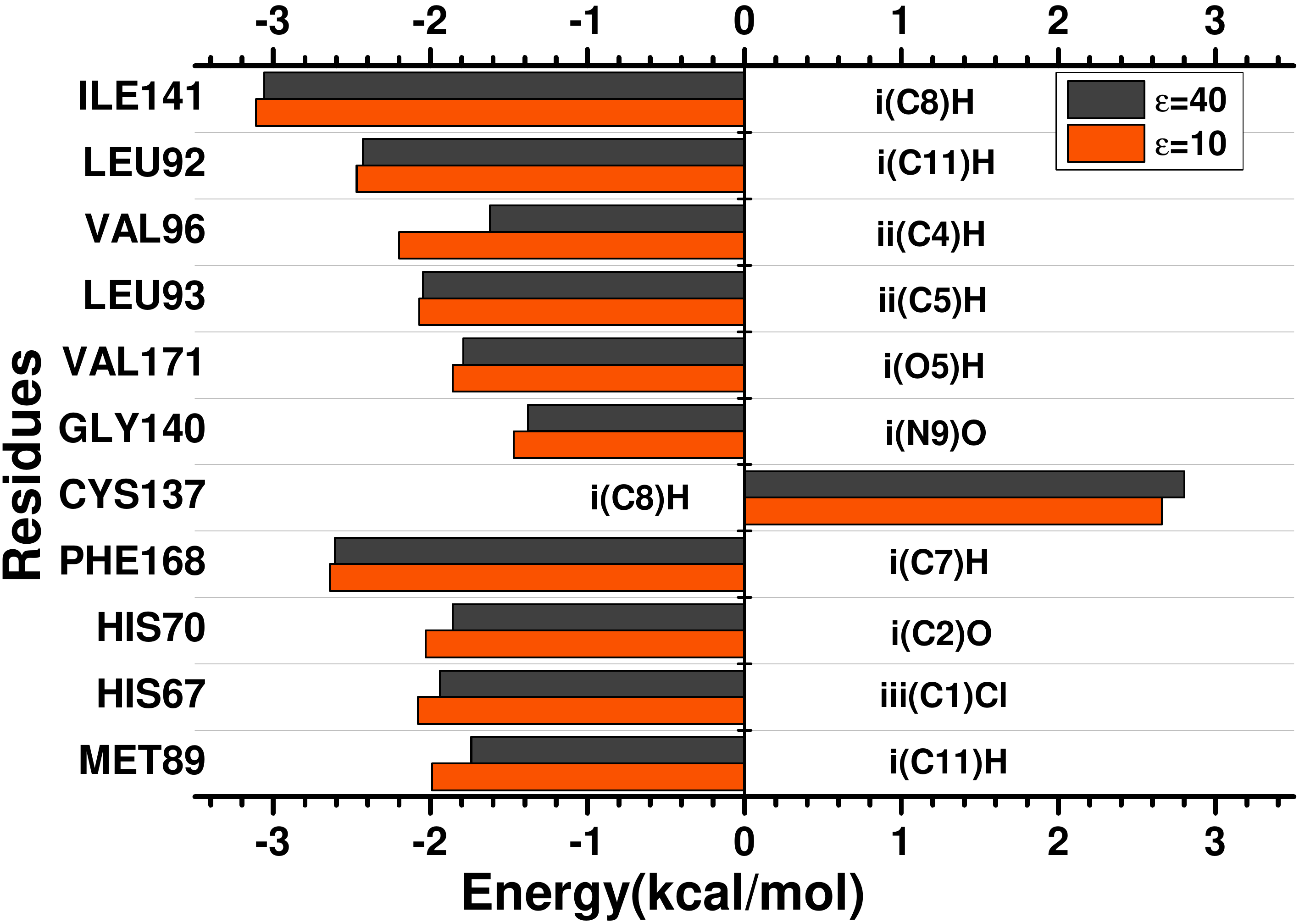}
\caption{The most relevant residues regarding the contribution of the interaction energies of the TtgR-CLM biological complex system. The regions and atoms of the ligands that interact with each residue at the binding site are also presented.}
\centering
\vspace{0.0cm}
\label{Fig7}
\end{figure}

The study of the interactions for the complex between the TtgR protein and the chloramphenicol (CLM) showed eleven relevant amino acid residues, where one amino acid, CYS137, presented a repulsive interaction with values of 2.66 kcal/mol ($\varepsilon$=10) and 2.80 kcal/mol ($\varepsilon$=40) and ten exhibited attractive energy values. Therefore, the attractive energy values in decreasing order are ($\varepsilon$=10; $\varepsilon$=40): ILE141 (-3.11; -3.06), PHE168 (-2.64; -2.61), LEU92 (-2.47; -2.43), VAL96 (-2.20, -1.62), HIS67 (-2.08; -1.94), LEU93 (-2.07; -2.05), HIS70 (-2.03; -1.86), MET89 (-1.99; -1.74), VAL171 (-1.86; -1.79) and GLY140 (-1.47; -1.38), in kcal/mol. Figure \ref{Fig7} shows the energy values achieved by the residue listed as the most important for the TtgR-CLM binding site, as well as the distance in angstroms ({\AA}) between residue-ligand.

The interaction energies by region were: for the region i -8.88 kcal/mol ($\varepsilon$=10) and -8.56 kcal/mol ($\varepsilon$=40); for region ii, -6.13 kcal/mol ($\varepsilon$=10) and -5.46 kcal/mol ($\varepsilon$=40); and for region iii, -4.11 kcal/mol ($\varepsilon$=10) and -3.8 kcal/mol ($\varepsilon$=40). Region i had the highest energy value since it has a greater amount of residue than other regions and contains those that achieved the highest attractive values. It is interesting to note that in this region, there are two charged atoms, one negatively (O9) and the other positively (N9). The ILE141 (Fig.\ 8) residue showed the highest energy value for the interaction between the TtgR-CLM complex, with attractive energies of -3.11 kcal/mol ($\varepsilon$=10) and -3.06 kcal/mol ($\varepsilon$=40) at a distance of 2.5 {\AA}, interacting with the effector by London forces in region i(C8)H. PHE168 reached values of -2.64 kcal/mol ($\varepsilon$=10) and -2.61 kcal/mol ($\varepsilon$=40) in 3.0 {\AA}, interacting with the effector in regions i(C7)H and ii(C5) by London and dipole-dipole-induced forces, respectively.

Meanwhile, the amino acid LEU92 demonstrated attractive energies of -2.47 kcal/mol ($\varepsilon$=10) and -2.43 kcal/mol ($\varepsilon$=40) interacting with CLM through London forces [i(C11)H] and dipole- dipole-induced [iii(C2)O]. Like the waste mentioned earlier, it is part of the nonpolar units that make up a large part of the connection pocket and is located on the {$\alpha$}5 helix. VAL96, on the other hand, presented the values of energies -2.20 kcal/mol ($\varepsilon$=10) and -1.62 kcal/mol ($\varepsilon$=40) in the radius 2.5 {\AA} connecting to the ligand by means of London forces [ii(C4)H]. This amino acid is located at {$\alpha$}5 and connects with the LEU66 residue, molding a hydrophobic bag for binding to chloramphenicol \cite{Daniels2010}.

The HIS67, which is arranged in the {$\alpha$}4 helix, showed attractive energy values of -2.08 kcal/mol ($\varepsilon$=10) and -1.94 kcal/mol ($\varepsilon$=40) at a minimum distance of 3.5 {\AA} demonstrating dipole-dipole forces in the region iii [iii(C1)Cl]. It was observed in another study that HIS67 is a residue that is probably involved in a portal where effectors can enter the general binding pouch, the same study found that an H67A mutation allowed the binding of the protein to naringenin, florentine, and chloramphenicol with affinities close to the wild type protein \cite{Daniels2010}. In another study, the importance of the HIS67, VAL96, CYS137, and VAL171 residues was also verified in a map of interactions with CLM, highlighting the stabilization effect promoted \cite{Sun2020}.

In turn, the LEU93 (Fig.\ 8) residue presented attractive energies of -2.07 kcal/mol ($\varepsilon$=10) and -2.05 kcal/mol ($\varepsilon$=40), interacting with the CLM through London ii (C5) H forces. It is located on the {$\alpha$}5 helix and contributes to the hydrophobic space of that helix. HIS70, on the other hand, indicated energy values of -2.03 kcal/mol ($\varepsilon$=10) and -1.86 kcal/mol ($\varepsilon$=40) in a radius of 3.0 {\AA}. It is positioned on the {$\alpha$}4 helix and interacts with the effector through region iii (C2) O through dipole-dipole-induced forces and region iii (C1) Cl through dipole-dipole connections. Like HIS67, this amino acid can also be involved in an effector entry portal in the TtgR protein binding pocket. The HIS70A mutation showed that TtgR was unable to bind to the CLM and was not released from the DNA binding region in the presence of this effector \cite{Daniels2010}.

MET89 is a residue that is located in the {$\alpha$}5 helix in a region close to the top of the pocket interactions and favors the hydrophobic environment of that location \cite{Alguel2007}. The energy values of -1.99 kcal/mol ($\varepsilon$=10) and -1.74 kcal/mol ($\varepsilon$=40) were observed within a radius of 3.5 {\AA}. This effector's interaction with the CLM occurred from London forces in region i(C11)H. Finally, VAL171 (Fig.\ 8) is located in the {$\alpha$}8 helix, and, like MET89, it provides the hydrophobic medium of the sidewalls of that helix. This expressed the energy values of -1.86 kcal/mol ($\varepsilon$=10) and -1.79 kcal/mol ($\varepsilon$=40) for a radius of 2.5 {\AA}, interacting with the effector by means of dipole-dipole-induced forces [ii(O5)H].

\begin{figure}[htb!]
\centering
\vspace{0.0cm}
\includegraphics[width=0.5\textwidth,keepaspectratio]{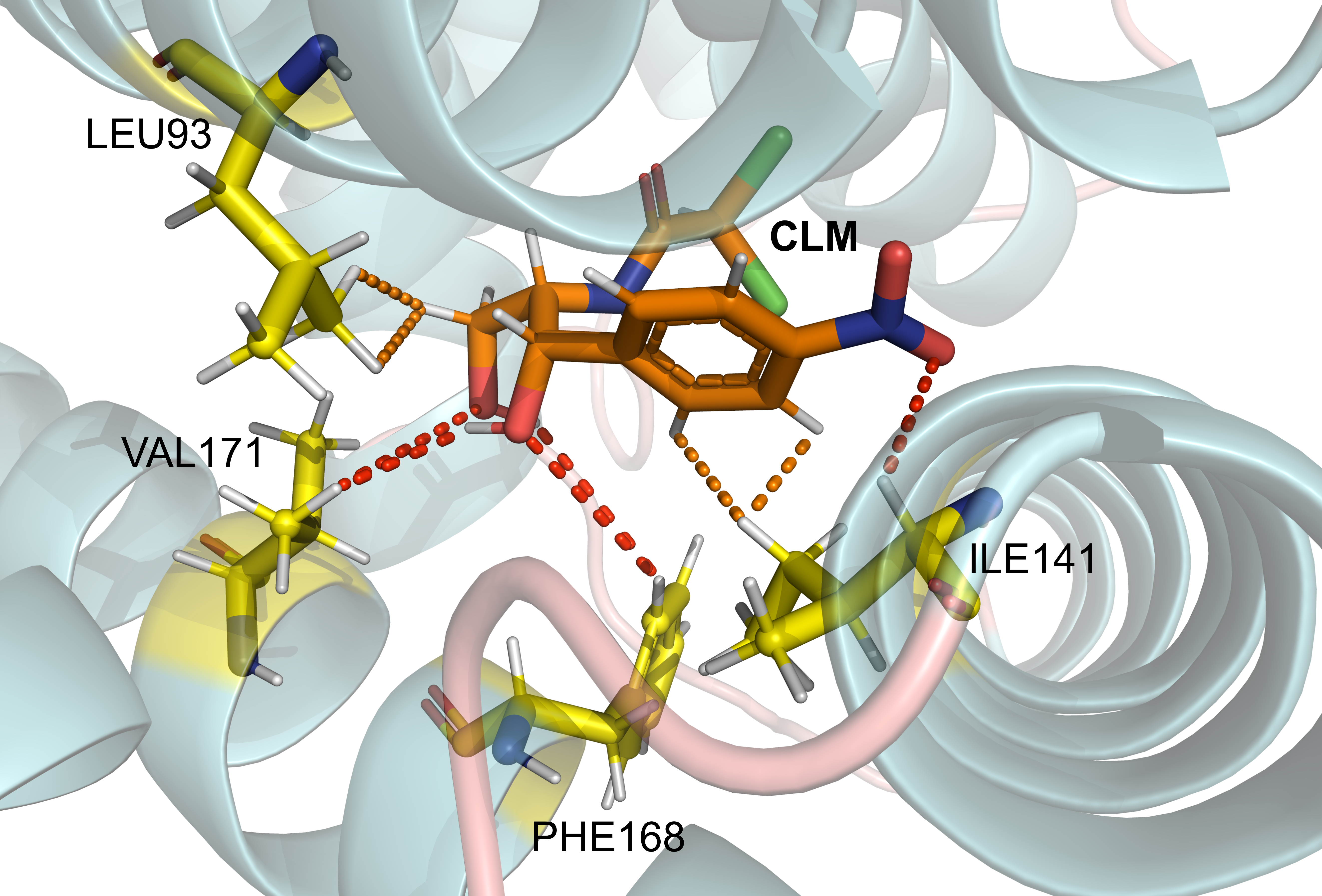}
\caption{The main interactions for the TtgR-QUE complex. Dashed lines correspond to non-conventional H bonds (red) and London forces (orange) interactions.}
\centering
\vspace{0.0cm}
\label{Fig8}
\end{figure}

Comparing the analysis of the energetic interactions between the TtgR-QUE, TtgR-TAC, and TtgR-CLM complexes, it was found that these effectors bind to the protein in a similar way. Furthermore, the interactions are predominantly controlled by hydrophobic residues that are part of the side walls of the helices that form this protein, such as LEU93, LEU92, ILE141, VAL171, and VAL96, and these interactions are relevant for protein-ligand recognition.

The amino acids VAL171, VAL96, ILE141, LEU92, and PHE168 were evaluated as energetically relevant for the three analyzed complexes. Residues HIS70, CYS137, and HIS67 were important for TtgR-TAC and TtgR-CLM systems, while MET167 and ASP172 were significant for TtgR-TAC and TtgR-QUE. LEU93 was considered relevant for TtgR-QUE and TtgR- CLM. On the other hand, the amino acids ARG176, GLU78, and LYS91 showed significant interaction energy only for TtgR-QUE. MET89 and GLY140 were relevant only for the TtgR-CLM complex. Finally, the TtgR-TAC complex, in a particular way, showed the residue ASN110 as significant.

\section{Conclusions}

In summary, this study went deeper into understanding the MDR resistance mechanism mediated by the multi-drug TtgR binding repressor of the microorganism \textit{P. putida} of the strain DOT-T1E, in complex with the antibiotics TAC and CLM and with the flavonoid QUE. This analysis aimed to investigate the conformations of these ligands to the binding site with this protein, presenting the main interactions between these complexes, to obtain strategies for designing new drugs against this resistance profile.

Through an energetic intermolecular calculation based on quantum chemistry, the most relevant residues of TtgR repressor responsible for recognition and formation of the binding pocket with QUE (TAC; CLM) effector were chemically and energetically described, namely: PHE168, VAL171, ILE141, MET167, LEU93, ARG176, LYS91, VAL96, and LEU92 (VAL96, ARG130, LEU113, VAL171, HIS70, ARG176, LEU92, ILE141 and LEU93; ILE141, PHE168, LEU92, VAL96, HIS67, LEU93, HIS70, MET89, VAL171, and GLY140).

According to the convergence study, the total binding energy of the ligand-protein systems followed the order TAC-TtgR > QUE-TtgR > CLM-TtgR. LEU93, LEU92, ILE141, VAL174, and VAL96 residues were relevant for the three complexes. ARG176 (HIS70; PHE168) presented a significant attractive character in TtgR-QUE and TtgR-TAC (TtgR-TAC and TtgR-CLM; TtgR-QUE and TtgR-CLM) complexes. Likewise, ASP172, GLU78, MET167, and LYS91 (GLU111, ASP64, ARG130, and LEU113; CYS137, HIS67, GLY140, and MET89) showed relevant interactions with TtgR-QUE (TtgR-TAC; TtgR-CLM) complex.

The computer simulations presented here might be valuable for a better understanding of the association of effectors to the TtgR repressor, which can overcome multiple drug resistance mechanisms.

\begin{acknowledgement}

This work was financed by the Coordenação de Aperfeiçoamento de Pessoal de Nível Superior (CAPES), Conselho Nacional de Desenvolvimento Cientifico e Tecnológico (CNPq), FAP-DF, and FAPESP. D.S.G thanks the Center for Computing in Engineering and Sciences at Unicamp for financial support through the FAPESP/CEPID Grants \#2013/08293-7 and \#2018/11352-7. L.A.R.J acknowledges the financial support from a FAP-DF grants $00193-00000857/2021-14$, $00193-00000853/2021-28$, and $00193-00000811/2021-97$, and CNPq grants $302922/2021-0$ and $350176/2022-1$. L.A.R.J. also gratefully acknowledges the support from ABIN grant 08/2019 and Fundaç\~ao de Apoio \`a Pesquisa (FUNAPE), Edital 02/2022 - Formul\'ario de Inscriç\~ao N.4.

\end{acknowledgement}






\bibliography{JCIM}

\end{document}